\documentclass[twocolumn,showpacs,preprintnumbers,prd,floatfix]{revtex4} 
\usepackage{epsfig}
\usepackage{array}
\usepackage{graphicx}
\usepackage{dcolumn}
\usepackage{bm}
\usepackage{epstopdf}
\usepackage{amssymb}
\usepackage{natbib}
\begin{document}
\title{Numerical Simulations of Oscillating Soliton Stars: Excited States in Spherical Symmetry and Ground State Evolutions in 3D}
\author{Jayashree Balakrishna$^{1}$, Ruxandra Bondarescu$^{2}$, Gregory Daues$^{3}$, Mihai Bondarescu$^{4,5}$}
\affiliation{$^{1}$Harris-Stowe State University, St.\ Louis, MO USA.\\
$^{2}$Cornell University, Ithaca, NY USA.\\
$^{3}$National Center for Supercomputing Applications, Urbana, IL  USA.\\
$^{4}$Max Planck Institut f\"ur Gravitationsphysik, Albert Einstein Institut, Golm, Germany.\\
$^{5}$California Institute of Technology, CA USA.\\
}
\date{\today}
\begin{abstract}
Excited state soliton stars are studied numerically for the first time. 
The stability of spherically symmetric $S$-branch excited state oscillatons under 
radial perturbations is investigated using a 1D code. We find that these stars 
are inherently unstable either migrating to the ground state or collapsing to 
black holes. Higher excited state configurations are observed to cascade 
through intermediate excited states during their migration to the ground state. 
This is similar to excited state boson stars \cite{jaya}. Ground state oscillatons 
are then studied in full 3D numerical relativity.   Finding the appropriate gauge 
condition for the dynamic oscillatons is much more challenging than in the case 
of boson stars. Different slicing conditions are explored, and a customized 
gauge condition that approximates polar slicing in spherical symmetry is implemented.  
Comparisons with 1D results and convergence tests are performed.  The behavior of 
these stars under small axisymmetric perturbations is studied and 
gravitational waveforms are extracted. We find that the gravitational waves damp
 out on a short timescale, enabling us to obtain the complete waveform.
This work is a starting point for the evolution of real scalar field 
systems with arbitrary symmetries. 
\end{abstract}
\maketitle
\section{Introduction}

 Real scalar fields play an important role in many models in particle physics 
and cosmology.  The axion is described by a real field as is the inflaton.
 Scalar particles are important dark matter candidates. Models of real scalar field dark matter have given
good fits for the velocity profile of spiral galaxies \cite{FSG2000}. Recently there has been widely
publicized observational evidence by NASA for the existence of dark matter with both the Hubble
telescope \cite{DarkRing} and the Chandra X-ray observatory \cite{BulletCluster}. These particles 
could come together through some kind of Jeans instability mechanism to form gravitationally 
bounded objects such as oscillating soliton stars (also called oscillatons) \cite{seidel94}. 

 Soliton stars possess a spectrum of states: ground state and excited states.   
The ground state scalar field has no nodes, a first excited state has one node 
and so on. The stability of spherically symmetric ground state oscillatons has been 
studied numerically \cite{seidel91, alcubierre02}.  In this paper we extend this 
study to the case of excited states. Studying the stability of excited states is 
important  because they may be intermediate states during the 
formation of these stars. 
 
 In this work we also begin an exploratory investigation of ground state soliton stars 
 in three-dimensional numerical relativity with the goal of performing
 stable evolutions and extracting gravitational waveforms. Numerical relativity has 
 made major progress in the past few years. It is now possible to evolve 
 binary black hole inspirals stably for many orbits \cite{Manuela,Caltech}. 
 Gravitational waveforms from black hole simulations 
have been extracted and match accurately against post-Newtonian estimates \cite{Post}.  
Gravitational wave detectors have reached their design sensitivity and waveform templates
are required to analyze data produced by the science runs \cite{ligo,LIGOUS}. Generating numerical
waveforms from other compact objects predicted by general relativity such as
soliton stars (comprised from real scalar fields)  and boson stars (made up from
complex scalar fields) is a timely and necessary effort.  Gravitational waveforms from two 
colliding boson stars \cite{thesis,TwoBSs,BSbinaries} as well as from single distorted 
boson stars \cite{us} have been extracted.  To our knowledge this is the first 
soliton star study in 3D numerical relativity.

Numerous authors have studied soliton stars numerically in spherical 
symmetry \cite{seidel91,alcubierre02,urena,InsideOscillatons}. Mathematically, oscillating 
soliton stars are nonsingular solutions to the Einstein-Klein-Gordon (EKG) equations 
represented by a massive real scalar field for which both the metric and the 
scalar field are periodic in time \cite{seidel91}.
The absence of static equilibrium configurations  contrasts oscillatons with the case of boson stars,
which are complex scalar field solutions to the EKG equations that have equilibrium
configurations characterized by static metric components (the scalar field has an
$\Phi(r,t) = \phi(r) e^{i\omega t}$ time dependence but the metric and energy itself
are time independent), and makes their evolution more challenging numerically.
 Similar to boson stars and neutron stars, the mass profile of ground state soliton 
 stars has a stable ($S$-branch) and an unstable branch ($U$-branch) \cite{seidel91,alcubierre02}. The mass 
profile has a maximum at  $M_{c\,\,\rm{ground\, st.}}=0.607 M_{\rm{Pl}}^2/m$  corresponding to 
a central density $\phi_1(0) = 0.48$ \cite{alcubierre02} (here $m$ is the mass of the scalar field.)
The configurations to the left (lower central density) of this critical mass are stable,
while those to the right (greater central density) are unstable. 

 In this paper we study the stability of excited state oscillatons under radial perturbations 
 and find that all excited configurations are inherently unstable. They either migrate 
 to the ground state or collapse to black holes. We present simulations for both scenarios for
configurations in the first excited state.  We find the mass profile of first excited state
configurations as a function of central density $\phi_1(0)$ (see  Fig.\ \ref{figmassexcited}.) It has a
maximum at a similiar critical density as for ground state
stars, but in this case it corresponds to a higher critical mass of 
$M_{c\,\,\rm{1st\, excited \, st.}} = 1.33 M_{\rm{Pl}}^2/m$.   We evolve  $S$-branch  one-node stars when 
no explicit perturbation is applied other than that induced by the numerical grid (we covered
the 95\% mass radius of the star with the same number of grid points 
$\sim 187$ to make meaningful comparisons) and find 
that stars that are unable  to migrate to the ground state collapse to black holes. 
The migration can occur only if the star can lose enough mass through scalar radiation 
to become an $S$-branch ground state configuration with $M < M_{c\,\,\rm{ground\,st.}} =
 0.607 M_{\rm{Pl}}^2/m$. A first excited state star of mass 
$M < 0.84 M_{\rm{Pl}}^2/m$ succeeds in losing its excess mass and migrates to 
the ground state. We study the relative times of collapse to black holes of 
different configurations for $S$-branch stars with mass $M > 0.84 M_{\rm{Pl}}^2/m$. 
The collapse times that we tabulate are approximate because polar slicing does not 
penetrate black hole horizons.  We find that 
this time decreases as the central density increases.  We also present a simulation 
of the migration of a one-node star to the ground state and find the ground state 
configuration that it migrated to.  We then simulate the migration of a 5-node 
configuration to the ground state. During the migration process the star cascades 
through a superposition of lower excited state configurations. We are able to determine 
an approximate 4-node intermediate state and then follow the evolution until 
the star settles into a ground state configuration. The simulations are performed 
using the polar slicing condition $K_\theta^\theta + K_\phi^\phi =0$, which is a 
natural choice for spherically symmetric spacetimes \cite{seidel90}.

In the second part of the paper we focus on 3D simulations of ground state soliton stars. We use the
Cactus Computational Toolkit \cite{cactus}  with the scalar field evolution of 
Guzman \cite{francisco} and the BSSN implementation \cite{stu,bssn} of the Einstein equations.  
Soliton stars retain some of the properties of boson star systems that made them a good testbed 
for numerical relativity in that the space-time is singularity free and the star
has a smooth outer boundary. However, the evolutions are challenging because the system is highly 
dynamic, i.e., the metric functions, extrinsic curvature components, and gauge 
functions are always rapidly oscillating in time.  In particular, finding an appropriate 
slicing condition and implementing it accurately to obtain
stable evolutions is a challenge for such dynamic spacetimes. 
We introduce a customized slicing condition based on the features of an eigenstate 
of a stable branch oscillaton in spherical symmetry    
and we refer to it as {\it truncated Fourier slicing}. 
Instead of enforcing that $K$ be zero as in the case of dynamic boson star 
evolutions \cite{us}, for this system 
we drive $K$ to the time dependent $ K^{\rm{(j_{max})}} (t) $, 
which is the trace of extrinsic curvature appropriate for a spherically 
symmetric oscillaton eigenstate expanded to finite order in
a Fourier series. While this condition is based on an eigenstate, we find that it is 
effective in simulations with small nonspherical perturbations applied to the star.
With this choice of gauge we are able to reproduce results from 1D codes within the 3D context,
and we are able to perform stable simulations of sufficient duration to extract gravitational waveforms.

We then study soliton stars under small nonradial perturbations and the emitted gravitational waves.  
We use two types of perturbations. The first type was  previously applied to boson stars 
by Guzman \cite{francisco}.
It consists of an imposed asymmetry in the grid resolution $\Delta x = \Delta y \ne \Delta z$,
while choosing the number of points $n_x = n_y \ne n_z$ such that  
the distance from the origin to the boundary of the grid is kept the same.
In this case the resolution itself is the perturbation. Consequently, the 
amplitude of the $\ell = 2, m = 0$ Zerilli waveform \cite{ZerRefs}  is expected to
be zero in the limit of infinite resolution and is observed numerically to
converge away with resolution at approximately second order. 
We next study an oscillaton under a perturbation  proportional to the $\ell=2, m=0$ spherical 
harmonic that perturbs the metric of the eigenstate nonradially. 
This perturbation is more physical as it could mimic a disturbance
in the gravitational field of the star due to the presence of another object. 
The Zerilli and Newman-Penrose $\Psi_4$ \cite{NPRefs} gravitational waveforms are extracted and 
compared at different detectors. The waveforms damp out on a short timescale. This
is consistent with the expectation that, similar to boson stars, 
soliton stars have only strongly damped modes because the scalar 
field extends to infinity and this allows energy to be 
radiated away rapidly \cite{futamase,kokkotas,toymodel}.  We also calculate the energy 
radiated in gravitational 
waves from the Zerilli function. By the end of our simulations the energy 
as a function of time is seen to flatten out, suggesting that the full 
gravitational waveform has been extracted. 

In Sec.\  \ref{equations} we describe the equations for spherically symmetric oscillatons. The eigenvalue problem and  boundary conditions are then discussed in Sec.\ \ref{boundary}.  In Sec.\ \ref{ExcitedResults} our results for the excited state evolutions are presented.  Sec.\  \ref{3Deq} discusses the evolution equations for the 3D code. The gauge conditions are presented next in Sec.\ \ref{slicing}  with convergence tests and comparison to 1D results. Sec.\ \ref{waveforms} details the application of small nonradial perturbations to ground state soliton stars and discusses the extracted gravitational waveforms for different perturbations.  The results are summarized in the conclusion. 

\section{Spherically Symmetric Configurations}
\subsection{Mathematical Background}
\label{equations}

The action describing a self-gravitating real scalar field in a 
curved spacetime is given by

\begin{eqnarray}
I = \int d^4 x \sqrt{-g} \left( \frac{1}{16 \pi G}R \, 
    -\frac{1}{2} [ g^{\mu \nu} 
    \partial_{\mu} \Phi  \, \partial_{\nu} \Phi  
    + V(\Phi) ]  \right), 
\label{action}
\end{eqnarray}

\noindent where $R$ is the Ricci scalar, $g_{\mu\nu}$ is the metric 
of the spacetime, $g$ is the determinant of the metric, $\Phi$ is the 
scalar field, $V$ its potential of self-interaction, and 
units with $\hbar=c=1$ have been used. Greek indices take values between 0 and 
3 and the Einstein summation convention is used. 
The variation of this action with respect to 
the scalar field leads to the Klein-Gordon equation, which can be written as

\begin{equation}
\Phi^{; \mu} {}_{;\mu} - \frac{1}{2} \frac{dV}{d\Phi}  = 0 . 
\label{kg-covariant}
\end{equation} 

\noindent When the variation of Eq.\ (1) is made with respect to the metric 
$g^{\mu\nu}$, the Einstein's equations $G_{\mu\nu}= 8\pi G T_{\mu\nu}$ 
arise, and the resulting stress energy tensor reads

\begin{equation}
T_{\mu \nu} = \partial_{\mu} \Phi \partial_{\nu}\Phi -\frac{1}{2}g_{\mu \nu}
[\Phi^{,\eta} \Phi_{,\eta} + V(\Phi))].
\label{setensor}
\end{equation}

\noindent In the present manuscript we focus on the free-field case, for 
which the potential is $V(\phi)=m^2 \Phi^2$, where $m$ is interpreted as 
the mass of the field.

The spherically symmetric line element is
\begin{equation}
ds^2=-N^2 dt^2 + g^2 dr^2 +r^2 (d\theta^2+ \sin^2\theta d\phi^2),
\end{equation}
where $N(r,t)$ is  the lapse function and $g^2(r,t)$ is the radial metric function.  We use a polar slicing condition and this ensures that the line element retains the above form throughout the 
evolution \cite{seidel91,alcubierre02}.  We follow Refs.\  \cite{alcubierre02,InsideOscillatons}  and write the coupled Einstein-Klein-Gordon equations as

\begin{eqnarray}
\label{AEq}
A' &= &4 \pi G r A ( C\dot{\Phi}^2 +\Phi'^2 +A m^2 \Phi^2)  + \frac{A}{r} (1-A), \\
\label{C1Eq}
C' &=& \frac{2C}{r} \left[ 1+A\left(4 \pi G r^2\Phi^2-1\right)\right], \\
\label{C2Eq}
C\ddot{\Phi} &=& -\frac{1}{2} \dot{C} \dot{\Phi} + \Phi''+\Phi' \left(\frac{2}{r} - \frac{C'}{2C}\right) - A m^2 \Phi, \\
\dot {A} &=& 8 \pi G r A \dot{\Phi} \Phi',
\label{Adot}
\end{eqnarray} 

\noindent where $A(r,t) = g^2$, $C(r,t) = [g(r,t)/N(t,r)]^2$ and $\Phi(r,t)$ is the real scalar field. 
These equations have no equilibrium solutions with static metric components \cite{seidel91}.
The simplest solutions are periodic expansions of the form

\begin{eqnarray} 
\Phi(t,r) &=& \sum_{j=1}^{j_{\rm max}} \phi_{2 j - 1}\cos((2j-1)\omega t), \\
\label{FourierS1}
A(t,r) &=& \sum_{j=0}^{j_{\rm max}} A_{2 j} (r) \cos(2j \omega t), \\ 
\label{FourierS2}
 C(t,r) &=& \sum_{j=0}^{j_{\rm max}} C_{2 j} (r) \cos(2j \omega t) ,
 \end{eqnarray}

\noindent  where $\omega$ is the fundamental frequency and $j_{\rm max}$ represents the value of $j$ at which the series is truncated for numerical computations. The actual solution is an infinite Fourier expansion of the above form that is convergent \cite{seidel91,alcubierre02}.
 
For numerical convenience in the code we use dimensionless variables
 \begin{equation}
  \Phi \to \sqrt{8 \pi G} \Phi,\quad  r \to r/m, \quad C \to C m^2/\omega^2, \quad t \to \omega t.
  \end{equation}

  \subsection{Boundary Conditions and Eigenvalue Problem}
  \label{boundary}
  
The system is asymptotically flat: $ A(r=\infty,t)=1$ and $\Phi(r=\infty,t) =0$. 
Thus the coefficients $A_0 (\infty)=1$, $A_j(\infty) = 0 $ for $ j \ne 0$, 
$\phi_j (\infty) =0$ for all $j$, and $C_j(\infty) = 0 $ for $ j \ne 0 $. 
The series is truncated at $j_{max}=2$ and the coefficients of $\sin(j t)$ 
and $\cos(j t)$ are matched in Eqs.\ (\ref{AEq}-\ref{C2Eq}). Eq.\ (\ref{Adot}) is used as an
algebraic relation to determine $A_0$ only.  The system of equations is solved as an 
eigenvalue problem for $C_j$ after specifying the central field $\phi_1(0)$. The lapse 
at the boundary gives the value of $\omega$. See Ref.\ \cite{InsideOscillatons} for 
further details. One sets $t=0$ in Eq.~(\ref{Adot}) to determine the initial  
metric components  $g_{rr}(t=0) = A_0+A_2 + A_4$, $g_{tt}(t=0) = -(A_0+A_2 + A_4)/(C_0+C_2+C_4)$.
    
  The configurations we consider are excited states. The fields $\phi_{2 j-1}$ with $ j=1,2,...$ have nodes.
  The first excited state field configurations have 1 node, the second have 2 and so on.
  
   After determining the initial configuration, the evolution of the system 
is studied using the 1D boson star evolution code of Refs.\ \cite{seidel90,jaya} 
with the field and its derivative having one component (real field) instead of 
two (complex field). The mass of the star is determined by the value of
$g_{rr} = A(t,r)$ at the edge of the grid
\begin{equation}
M = \lim_{r\to\infty} \frac{r}{2} \left[1 - \frac{1}{A(t,r)}\right] \frac{M_{\rm{Pl}}^2}{m}.
\label{masseq}
\end{equation} 

\subsection{Stability of Excited States}
\label{ExcitedResults}
 
\begin{figure}
\begin{center}
\leavevmode
\epsfxsize=250pt
\epsfbox{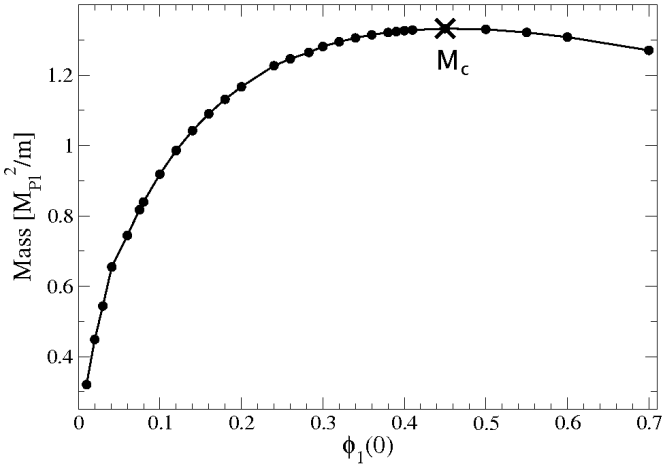}
\caption{The mass is shown as a function of central density for stars in their first excited state. 
The mass increases monotonically until it reaches a critical point  at $\phi_{1c} \approx 0.45$, 
and $M_c \approx 1.33 M_{\rm Pl}^2/m$ after which it starts decreasing. This point marks the end 
of the $S$-branch, and the beginning of the $U$-branch. These stars are more 
massive than ground state stars.}
 \renewcommand{\arraystretch}{0.75}
 \renewcommand{\topfraction}{0.6}
  \label{figmassexcited}
\end{center} 
\end{figure}

\begin{figure}
\begin{center}
\leavevmode
\epsfxsize=250pt
\epsfbox{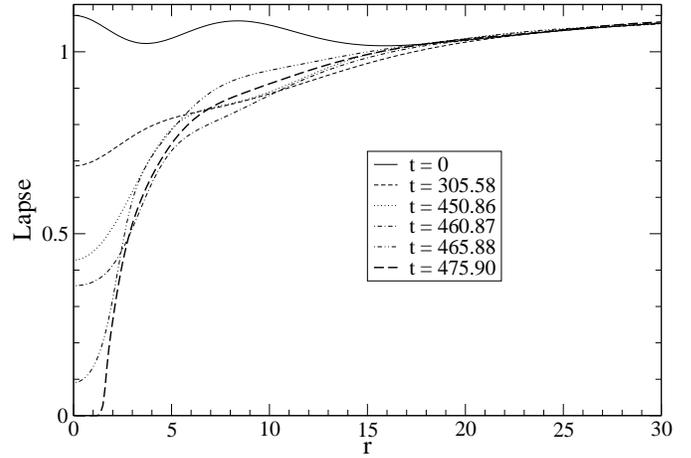}
\newline\newline\newline\newline
\epsfxsize=250pt
\epsfbox{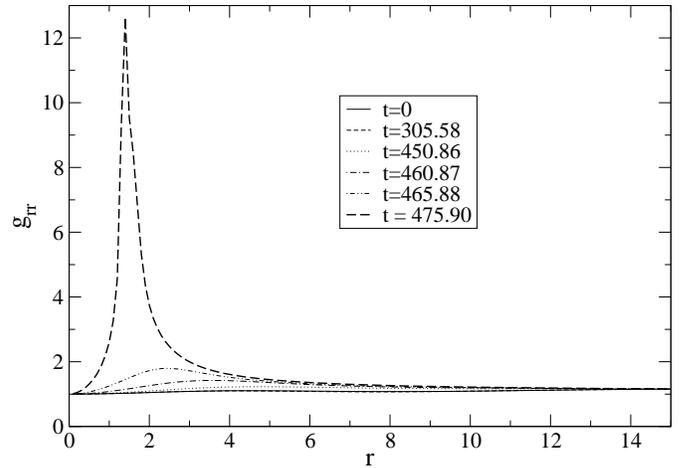}
\caption{(a) The lapse and (b) the radial metric are shown at different times as a star with 
a central density of $\phi_1(0) = 0.2828$ collapses to a black hole. The polar slicing 
condition causes the lapse to collapse and the metric to rise sharply 
as an apparent horizon is approached.}
 \renewcommand{\arraystretch}{0.75}
 \renewcommand{\topfraction}{0.6}
\label{blackhole1}
\end{center}
\end{figure}

\begin{figure}
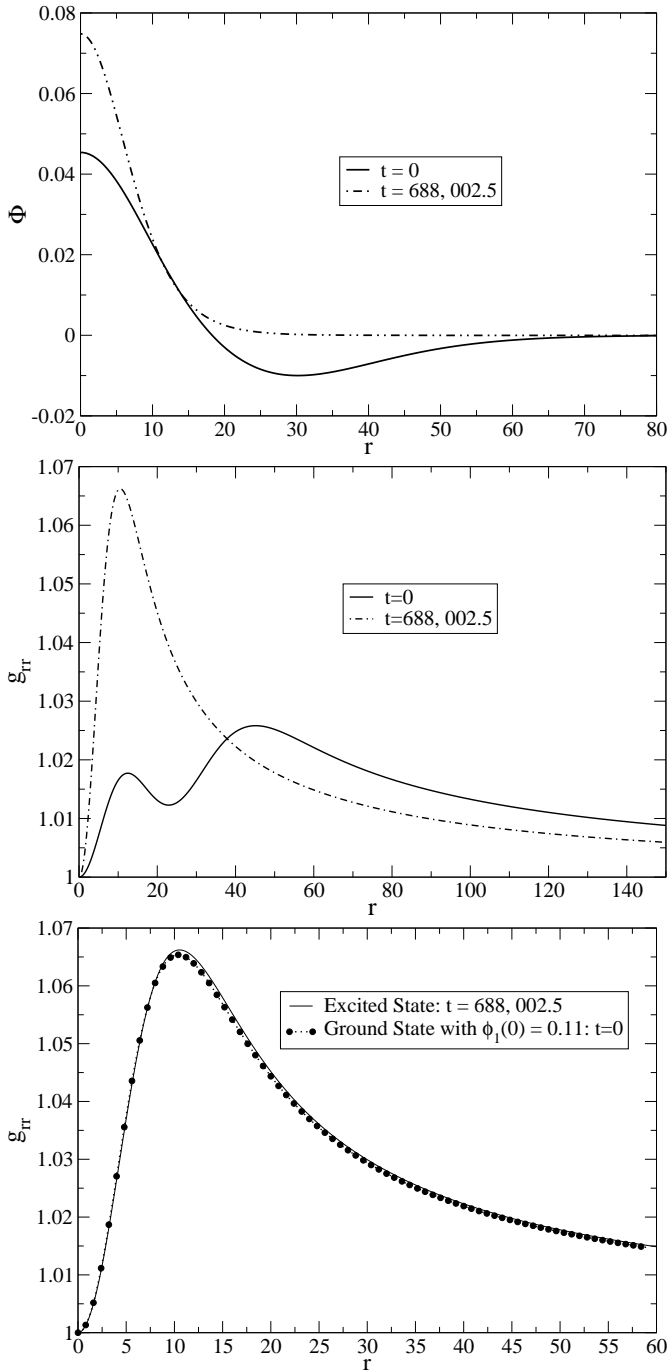

\begin{center}
\leavevmode
\epsfxsize=250pt
\epsfbox{Fig3a.eps}
\newline
\epsfxsize=250pt
\epsfbox{Fig3b.eps}
\newline
\epsfxsize=250pt
\epsfbox{Fig3c.eps}
\caption{ In (a) the scalar field $\Phi$ and (b) metric component $g_{rr}$ are plotted against radius
at $t=0$ and $t \approx 690, 000/m$. The initial $g_{rr}$ has two maxima (first excited state star; 
$\phi_1(0) = 0.041$ and $R_{95} = 51.35/m$).  By $t\approx 690, 000/m$ 
the star has migrated to a denser ground state configuration of $R_{95} \approx 16/m$ with lower mass.
(c)  The final metric $g_{rr}$ is compared to that of a ground state $t=0$, 
$\phi_1(0) = 0.11$ star. Although the ground state $g_{rr}$ in slightly lower 
because the star has not fully settled, the overall agreement is very good.}
 \renewcommand{\arraystretch}{0.75}
 \renewcommand{\topfraction}{0.6}
\label{migration1}
\end{center}
\end{figure}

\begin{figure}
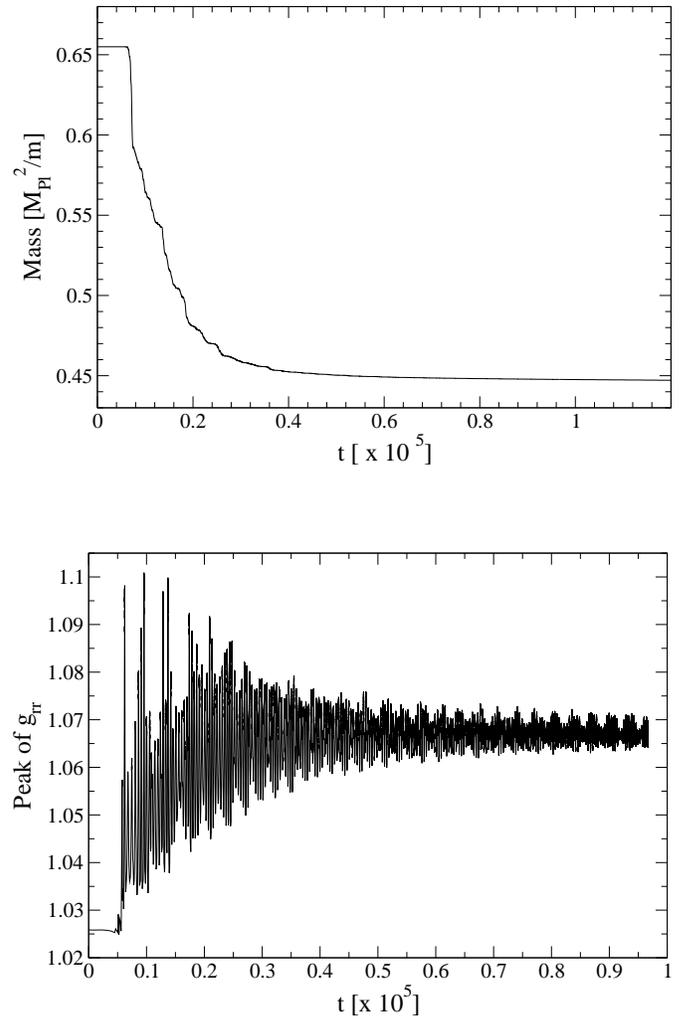

\begin{center}
\leavevmode
\epsfxsize=250pt
\epsfbox{Fig4a.eps}
\newline\newline\newline\newline
\epsfxsize=250pt
\epsfbox{Fig4b.eps}
\caption{(a) The mass of the same $\phi_1(0) = 0.041$ star is shown as the star migrates to 
the ground state. The mass flattens out as the star is settling towards a final ground 
state configuration. The oscillations in the metric also settle down to an approximately 
constant oscillation. The mass of the star at the end of the run is  $M \approx 0.44 M_{\rm Pl}^2/m$, 
and is within 2\% of the ground state configuration  ($\phi_1(0) = 0.11$ and $M=0.43 M_{\rm Pl}^2/m$) that 
the star is estimated to migrate to. (b) The maximum of $g_{rr}$ is shown as a function of time. 
The star loses more than 30\% of its mass during the migration.}
 \renewcommand{\arraystretch}{0.75}
 \renewcommand{\topfraction}{0.6}
\label{migration2}
\end{center}
\end{figure}

\begin{figure}
\begin{center}
\leavevmode
\epsfxsize=250pt
\epsfbox{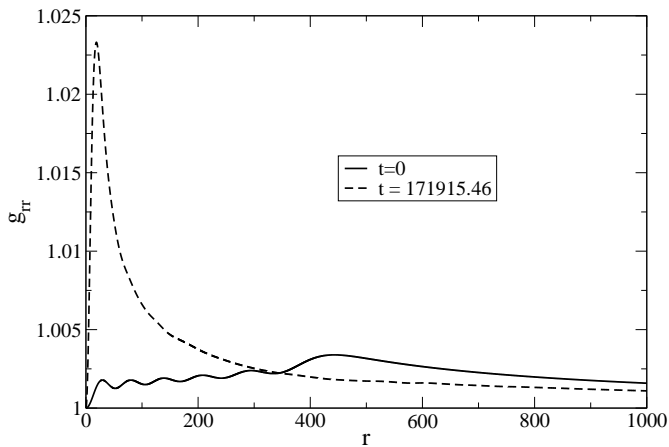}
\newline\newline\newline\newline
\epsfxsize=250pt
\epsfbox{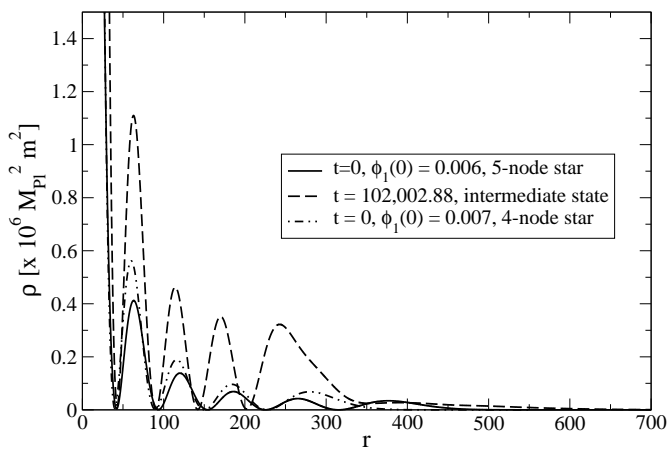}
\caption{(a) The radial metric $g_{rr}$ of a migrating five node soliton star configuration 
($\phi_1(0) = 0.006$; $M=0.79 M_{\rm Pl}^2/m$) is displayed as a function of 
radius on two timeslices. The $t=0$ radial metric has 6 maxima, which is characteristic to 
5th excited state configurations.  By $t\approx 172,000/m$ the star has migrated to the ground 
state. Although, the migration process is not yet complete by this stage, the star has 
lost 29\% of its mass through scalar radiation and the radial metric has 
only one maximum. (b) The central density of the same star is shown at $t=0$ and in a intermediate 
state at $t\approx 102,000/m$. In this intermediate state the star has four nodes (central density 
has five maxima). We plot the density of a four node star with $\phi_1(0) = 0.007$ for comparison.}
 \renewcommand{\arraystretch}{0.75}
 \renewcommand{\topfraction}{0.6}
  \label{fig5Nodes}
\end{center} 
\end{figure}
The mass profile of ground state soliton stars have been described in 
Refs. \cite{InsideOscillatons, alcubierre02}. In Fig. \ref{figmassexcited} we show the 
mass of first excited state oscillatons as a function of $\phi_1(0)$.
 These stars have two branches similar to the ground state configurations with the  mass 
increasing to a maximum mass of  $M_c \approx 1.3 M_{\rm Pl}^2/m$. This is higher than the maximum 
mass of a ground  state star.  In general, excited configurations of a given central field density $\phi_1(0)$ are larger and more massive than ground state configurations with the same central density.
The branch to the left of the maximum is traditionally called the $S$-branch and the branch 
to the right is called the $U$-branch. 
 
In the case of ground state stars the $S$-branch is found to be stable to radial perturbations \cite{alcubierre02}. 
Stability here means that stars on this branch move to new lower mass configurations on the same 
branch under small perturbations. $U$-branch stars are inherently unstable to small perturbations and
collapse to black holes.  Under large perturbations that reduce their mass sufficiently
they can migrate to the $S$-branch \cite{alcubierre02}.

In this work we study the stability of excited state $S$-branch oscillaton stars. In Fig.\ \ref{blackhole1}  
the evolution of the radial metric $g_{rr}$ of an $S$-branch first excited state star is shown. 
The central density of this configuration is $\phi_1(0)=0.2828$ and no explicit 
perturbation has been applied. 
The numerical grid ($\Delta r=0.1$) is the only source of perturbation.
As can be seen from the sharp rise in the radial metric (Fig.\ \ref{blackhole1}(a)) 
and collapse of the lapse (Fig.\  \ref{blackhole1}(b)) the star is going to form a 
black hole. The polar slicing condition causes the lapse to collapse and the radial 
metric to rise sharply when an apparent  horizon is approached \cite{seidel90}.  
Clearly, the star is unstable although this configuration lies on the $S$-branch. 


We next study the time of collapse of one node $S$-branch oscillaton configurations to black 
holes as a function of central density.
Excited states are inherently unstable and the discretization error due to the numerical grid is 
sufficient to make them collapse to black holes or migrate to the ground state. To compare the 
effects of radial perturbations on stars with different central densities, 
we use the discretization of the grid as a perturbation with the stars covered by 
the same numbers of grid points. The stars need to be resolved differently 
because of their different central densities.
In the following simulations the 95\% mass radius $R_{95}$ was considered to be the radius of the star.

The first configuration considered is close 
to the critical mass and has $\phi_1(0)  = 0.41$ and a radius of 
$R_{95} = 14.9/m$.  This was covered by 187 grid points with a resolution $\Delta r \approx 0.08$.
The second configuration has $\phi_1(0)=0.2828$ star with  $R_{95}=18.7/m$ and was covered by the
same number of grid points with $\Delta r = 0.1$.  This star collapsed in a longer time. We took the 
time of collapse (polar slicing does not allow us to find the actual time of formation of a 
black hole horizon \cite{seidel90}) to be the beginning of the sharp rise in the metric. 
Coordinate grid stretching \cite{any_bh_paper} (due to the differential rates that test 
particles on geodesic curves fall into the black hole) causes the metric function $g_{rr}$ to rise 
and form a sharp peak.  The formation of the peak signals the approach of horizon 
formation on a short time scale.

In Table \ref{tablecollapse} we show the time of collapse to a black hole  for different 
$S$-branch configurations. The table shows the central field density, mass, 
and radius of each configuration. The time of collapse to a black hole increases as 
the central density decreases until a central density of about $\phi_1(0) = 0.075$.  For central 
densities of $0.075$ and lower the stars did not collapse but migrated to configurations in the ground state. 
Although a first excited state star of central density $\phi_1 = 0.075$ is more massive than any ground 
state star (maximum mass of ground state configurations is about $0.61 M_{\rm Pl}^2/m$) it lost the 
excess mass through scalar radiation and settled in the ground state.
Excited state $S$-branch stars are clearly inherently unstable and either 
collapse to black holes or migrate to the ground state. 
\begin{table}[h]
\begin{center}
\footnotesize
\begin{tabular}{|cccc|} \hline
\multicolumn{4}{|c|}{$S$-branch Soliton Stars collapse times} \\
\hline
 $\phi_1(0)$ & M($M_{\rm Pl}^2/m$)& $R_{95}$(1/m)&  collapse time  \\ \hline
\hline
 0.41&1.32& 14.9 &   250      \\
0.2828&1.26& 18.7& 445         \\
0.1414&1.07&27.0 &1280         \\
 0.10&0.92&33.7& 2250 \\ 
0.08&0.84&38.0&4160  \\        
 0.075&0.82&39.5&goes to gr. state \\  \hline
\end{tabular}
\end{center}
\caption{Physical characteristics of $S$-branch soliton stars in the first excited state are listed together with the collapse time to black holes. Each star is covered by the same number of grid points 
 ($\sim 187$) to facilitate the comparison. The radius of the star is taken to be the radius within which $95\%$ of the mass of the star is contained. Configurations with a mass below $0.84 M_{\rm Pl}^2/m$ lose enough mass through scalar radiation to migrate to the ground state.}
\label{tablecollapse}
\end{table}

In Fig.\ \ref{migration1} and Fig.\ \ref{migration2} we show a first excited state star with 
$\phi_1(0)=0.041$ of mass $0.655 M_{\rm Pl}^2/m$ migrating to the ground state. 
No explicit perturbation was applied.  
In Fig.\ \ref{migration1}(a) the initial (one node) and final (no nodes) field configurations are displayed. 
In Fig.\ \ref{migration1}(b) the initial and final radial metric $g_{rr}$ are shown.
The initial radial metric has two maxima, 
which is characteristic of a first excited state configuration.  The final radial metric 
($t=218, 998 \pi/m$) has one maximum indicating the star has gone to the ground state. 
The final configuration shown corresponds to a time which is an integral multiple of $\pi$. 
This is to ensure that we get the same phase as the $t=0$ configuration to facilitate 
comparison. In Fig.\ \ref{migration1}(c) the radial metric of the star at the end 
of the run is plotted together with  the radial metric of a ground state 
star of central density $\phi_1(0)=0.11$. The final radial metric of the migrating 
star is very close, to this ground state configuration. For comparison the mass 
of the star at the end of the run was about $M = 0.44 M_{\rm Pl}^2/m$ as compared to 
a mass of $M=0.43 M_{\rm Pl}^2/m$ for a $\phi_1(0)=0.11$ star. The migrating star 
is settling to a configuration close to the $\phi_1(0)=0.11$ ground state star. 

The mass loss is calculated using Eq.\ (\ref{masseq}) for the above 
configuration and is plotted as a function of time in 
Fig.\ \ref{migration2}(a). The star loses more than $30\%$ of its mass 
through scalar radiation.  The amount of mass loss decreases in time as the star 
settles to its final ground state configuration.  
In Fig.\ \ref{migration2}(b) the maximum of the metric function $g_{rr}$ of a perturbed 
excited state star is plotted against time.  As the star settles into its new ground state
configuration, it oscillates at two different frequencies. One is approximately the fundamantal unperturbed 
oscillation of period $\pi/\omega$ ($t_{\rm{code}} = \pi$). The second has much higher 
period that changes as the star goes through different intermediate configurations 
and eventually settles into the quasinormal mode of the ground state 
configuration \cite{alcubierre02}.

We next simulate the migration of an oscillaton in the 5th excited state to the ground state.
The initial configuration is that of a five node excited state $S$-branch star of central 
density $\phi_1(0)=0.006$ and  mass $M=0.792 M_{\rm Pl}^2/m$. As usual, no explicit has been applied
perturbation other than that due to the discretization of the numerical grid 
(resolution $\Delta r = 0.1$).  In Fig.~\ref{fig5Nodes}(a)  we show the radial metric $g_{rr}$ of the star 
at $t=0$, and $t\approx 172, 000/m$. The initial configuration ($t = 0$) has six maxima in the radial metric. 
By the end of the simulation the radial metric has just one maximum indicating the star has 
gone to the ground state (the mass at this stage is $M= 0.463 M_{\rm Pl}^2/m$.) During the evolution the star appears 
to pass through an intermediate four node state. This can be seen in Fig.~\ref{fig5Nodes}(b). 
In the figure $\rho/\cos(2t)$ ($\rho$ is the density of the star and the division by $\cos(2t)$ 
is in order to facilitate comparison as it helps  avoid phase differences) is plotted at 
$t=0$ and $t\approx 102, 000/m$ against the $t=0$ density of a four node $\phi_1(0)=0.007$ star.  
The evolving star initially had a density $\rho$ with six maxima characteristic of a five node star. 
By time $t=102, 000/m$ it has lost one of these maxima and appears to be in an intermediate 
four node state. We have plotted the density of a four node $\phi_1(0)=0.007$ star to compare 
the positions of the peaks and profile of $\rho$. The positions of the peaks are close to each 
other (especially the first three) but the sizes of the peaks are not. The star probably cascades 
through superpositions of lower excited configurations due to the inherent 
non-linearity of the system.

\section{Numerical Simulations of Ground State Soliton Stars in 3D}
\label{3D}


\subsection{3D Evolution equations}
\label{3Deq}

In order to find solutions to the Einstein-Klein-Gordon 
system of equations we use the 3+1 decomposition of Einstein's equations, 
for which the line element can be written as

\begin{equation}
   d s^{2}  = - \alpha^2  dt^2 + \gamma_{ab}  (dx^a + \beta^{a} dt) (dx^b 
              + \beta^b dt),  
\label{lineelem}
\end{equation}

\noindent
where $\gamma_{ab}$ is the three dimensional metric; from now on latin 
indices label the three spatial coordinates. The functions $ \alpha $ 
and $ \beta^{a}$ in Eq. (\ref{lineelem}) are freely specifiable gauge 
parameters, known as the lapse function and the shift vector respectively,
and $\gamma$ is the determinant of the 3-mentric.
Throughout the rest of the paper the standard general 
relativity notation is used. The Greek indices run from 0 to 3 and the 
Latin indices run from 1 to 3. In this section we use geometric units with
$G = c = 1$.

The Klein-Gordon equation can be written as a first-order evolution system 
by defining two new variables in terms of combinations of their derivatives: 
 $ \pi= (\sqrt{\gamma}/\alpha) 
(\partial_t \phi - \beta^c \partial_c \phi) $ and  
$\psi_{a}=\partial_a \phi$. With this 
notation the evolution equations become 

\begin{eqnarray}
 \partial_t \phi &=&  \frac{\alpha}{\gamma^{\frac{1}{2}}} \pi + 
\beta^a \psi_{a}, \\\nonumber
 \partial_t \psi_{a} &=& \partial_a( \frac{\alpha}{\gamma^{\frac{1}{2}}} 
\pi + \beta^b \psi_{b}), \\\nonumber
 \partial_t \pi &=& \partial_a (\alpha \sqrt{\gamma} \phi^a) 
  - \frac{1}{4} \alpha  \sqrt{\gamma} \frac{\partial V}{\partial 
  \Phi}. \label{kg-equations}
\end{eqnarray} 

\noindent
On the other hand, the geometry of the spacetime is evolved 
using the BSSN formulation of the 3+1 decomposition. According to this 
formulation, the variables to be evolved are 
$\Psi = \ln\gamma/12$, 
$\tilde{\gamma}_{ab} = e^{-4\Psi}\gamma_{ab}$, $K = \gamma^{ab}K_{ab}$, 
$\tilde{A}_{ab}=e^{-4\Psi}(K_{ab}-\gamma_{ab} K/3)$ and 
the contracted Christoffel symbols 
$\tilde{\Gamma}^{a}=\tilde{\gamma}^{bc}\Gamma^{a}_{bc}$,
instead of the usual ADM variables $\gamma_{ab}$ and $K_{ab}$. The 
evolution equations for these new variables are described in Refs. \cite{bssn, 
stu}:

\begin{eqnarray}
\partial_t \Psi &=& - \frac{1}{6} \alpha K \label{BSSN-MoL/eq:evolphi}, \\
\partial_t \tilde{\gamma}_{ab} &=& - 2 \alpha \tilde{A}_{ab},
\label{BSSN-MoL/eq:evolg} \\
\partial_t K &=& - \gamma^{ab} D_a D_b \alpha  \nonumber  \\ 
             &+& \alpha \left[
        \tilde{A}_{ab} \tilde{A}^{ab} + \frac{1}{3} K^2 + \frac{1}{2}
        \left( -T^{t}{}_{t} + T \right) \right],
\label{BSSN-MoL/eq:evolK}   \\
\partial_t \tilde{A}_{ab} &=& e^{-4 \Psi} \left[
 - D_a D_b \alpha + \alpha \left( R_{ab} - T_{ab} \right) \right]^{TF}
                    \nonumber  \\
        && + \alpha \left( K \tilde{A}_{ab} - 2 \tilde{A}_{ac}
\tilde{A}_b^c
        \right), \label{BSSN-MoL/eq:evolA}\\
\frac{\partial}{\partial t} \tilde \Gamma^a
&=& - 2 \tilde A^{ab} \alpha_{,b} + 2 \alpha \Big(
\tilde \Gamma^a_{bc} \tilde A^{cb}                              \nonumber \\
&& - \frac{2}{3} \tilde \gamma^{ab} K_{,b}
- \tilde \gamma^{ab} T_{b t} + 6 \tilde A^{ab} \phi_{,b} \Big)
                                                                \nonumber \\
&& - \frac{\partial}{\partial x^b} \Big(
\beta^c \tilde \gamma^{ab}_{~~,c}
- 2 \tilde \gamma^{c(b} \beta^{a)}_{~,c}
+ \frac{2}{3} \tilde \gamma^{ab} \beta^c_{~,c} \Big), 
\label{BSSN-MoL/eq:evolGamma2}
\end{eqnarray}

\noindent where $D_a$ is the covariant derivative in the spatial
hypersurface, $T$ is the trace of the stress-energy 
tensor~(\ref{setensor}) and the label $TF$ indicates the trace-free part 
of the quantity in brackets.
The coupling between the evolution of the BSSN variables and 
the variables describing the evolution of the scalar field is first order. 
That is, Eqs.\ (5) are solved using the method of lines with a 
modified version of the second order iterative Crank-Nicholson (ICN) 
integrator (see Ref.\ \cite{urena2}). After a full time step the stress-energy 
tensor in Eq.\ (3) is calculated and used to solve the BSSN evolution equations 
with an independent evolution loop based on the standard second order ICN 
\cite{icn}. 

In order to set up the correct scaled quantities to be evolved we use dimensionless 
variables. For an equilibrium configuration the real scalar field can be expanded in 
a Fourier series of the form $\Phi(r,t)=\sum_{j=1}^{j_{\rm{max}}} 
\phi_{2 j - 1} (r) \cos((2j-1)\omega t)$, where the total central density is 
$\phi(0) = \sum_{j=1}^{j_{\rm{max}}} \phi_{2 j - 1}(0)$. This implies that the stress 
energy components given in Eq.\ (3) also have a periodic time dependence.  The 
characteristic frequency $\omega$ together with the mass of the boson $m$ 
provide the natural dimensionless units, 

\begin{eqnarray}
r_{\rm{code}} &=&m r/M_{\rm{Pl}}^2, \qquad t_{\rm{code}} = \omega t, 
  \\\nonumber
\phi_{\rm{code}} &=&\frac{\sqrt{4 \pi}}{M_{\rm{Pl}}} \phi,\quad 
   \alpha_{\rm{code}} = \alpha \frac{m}{M_{\rm Pl}^2 \omega},
\end{eqnarray} 

\noindent which are the ones used within the present analysis. For 
further code details refer to Ref.\ \cite{francisco}.

\subsection{Slicing Condition and Convergence Tests}
\label{slicing}

In this section we introduce the slicing condition used for our 3D simulations. 
We base it on the observed dynamics of the trace of the extrinsic curvature tensor $ K $ for 
oscillaton eigenstates, and find that 
the slicing is accurate for evolving spherically symmetric oscillatons with small nonradial perturbations.
For soliton star eigenstates $K (t=0) = K_{rr}(t=0) = 0 $.
However, as time evolves the metric function $g_{rr}$ 
changes with time and consequently K oscillates periodically, 
given by
\begin{eqnarray}
 K &=&  \frac{   \gamma^{ij} {\dot \gamma}_{ij}  }{ -2\alpha }   =
 \frac{   \gamma^{rr} {\dot \gamma}_{rr}  }{ -2\alpha }   = 
 \sum_{j=1}^{\infty} K_{2 j} \sin(2 j t).
\label{K1}
\end{eqnarray}
We denote K expanded to order  $j_{\rm{max}}$ as 
\begin{eqnarray}
 K^{ (j_{\rm max}) } \equiv \sum_{j=1}^{j_{\rm{max}}} K_{2 j} \sin(2 j t).
\label{orderK}
\end{eqnarray}
In the case of spherical symmetry with the gauge condition provided by polar slicing 
 one can determine $K_{2j}$ by using the Fourier expansion of 
$\gamma_{rr}$ and $\alpha = \sqrt{A/C}$ from Eqs.\ (\ref{FourierS1}-\ref{FourierS2}) in the expression 
for $K$ (Eq.\ (\ref{K1})).  One can match the coefficients of 
$ \sin(2 j t) $ for each $ j $ 
and obtain $K_{2 j}$ in terms of a combination of $A_{2 j}$ and $C_{2 j}$.
The first two coefficients $K_2$ and $K_4$ are approximately given by
\begin{equation}
K_2 \approx A_2 \left( \frac{C_0}{A_0^3}  \right)^{1/2},  \label{kay2}
\end{equation}
\begin{equation}
K_4 \approx \frac{ A_2 C_2 }{ 4 C_0^{1/2} A_0^{3/2} } + \frac{2 A_4 C_0^{1/2} }{ A_0^{3/2} } - 
      \frac{3 A_2^2  C_0^{1/2} }{ 4 A_0^{5/2} }.   \label{kay4} 
\end{equation}
For 3D simulations we propose to enforce that $K$ take the value $ K^{(j_{\rm max})} $ 
of Eq.\ (\ref{orderK}), and do so in stable manner.
Various authors\cite{Kdriver} have noted that one can enforce 
maximal slicing ${\dot K} = 0 = K $ in a stable manner by solving 
\begin{equation}
\frac{\partial K}{\partial t} = - c K, 
\end{equation}
where $c$ is a positive constant.
Analogously, we drive $ K $ to the expanded $ K^{ ( j_{\rm max} ) }  $ 
for an eigenstate using
\begin{equation}
   \frac{\partial}{\partial t}  \left( K - K^{ ( j_{\rm max} ) } \right) = - c \left( K - K^{ ( j_{\rm max} ) } \right).  
\label{driver}
\end{equation}
Using the identity 
\begin{equation}
\label{dK}
      \nabla^2 \alpha - K_{ab}  K^{ab} \alpha - 4 \pi (S + \rho) \alpha = - \frac{\partial K}{\partial t} , 
\end{equation}
where $ \rho $ is the energy density and $ S $ is the trace of the spatial stress tensor, 
we rewrite Eq.\ (\ref{driver}) as an elliptic equation to solve for the lapse 
\begin{eqnarray}
      \nabla^2 \alpha - K_{ab}  K^{ab} \alpha - 4 \pi (S + \rho) \alpha 
      &=& - \frac{\partial K^{ ( j_{\rm max} ) }}{\partial t} \\ \nonumber
&+& c \left( K - K^{ ( j_{\rm max} ) } \right).  
\label{ellip}
\end{eqnarray}
We refer to the slicing obtained from this gauge condition (Eqs.\ (\ref{driver}) 
and (\ref{ellip})) as {\it truncated Fourier slicing}.

For the 3D simulations described in this section and the subsequent section, we use a ground state oscillating soliton star with
\begin{eqnarray}
\phi_1(0) = 0.20,\;\omega = 0.91 \frac{m}{M_{\rm Pl}^2}, \\ \nonumber
M = 0.57 \frac{M_{\rm Pl}^2}{m},\; R_{95} = 8.8 \frac{M_{\rm Pl}^2}{m},
\end{eqnarray} 
and we label this as the Model $S_1$ oscillaton for future reference. For initial data within the 3D simulations we take high resolution 1D spherical data for this model and interpolate it onto the 3D Cartesian grid using bspline interpolation. All our 3D simulations are performed in octant symmetry.

\begin{figure}
\begin{center}
\leavevmode
\epsfxsize=250pt
\epsfbox{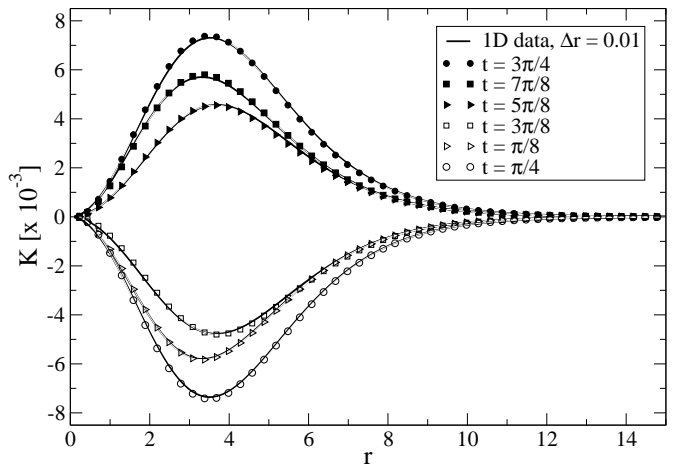}
\caption{The extrinsic curvature $K$ is shown for a Model $S_1$ star on different timeslices 
as a function of $r$.  The solid lines represent the results obtained from 
a 1D spherically symmetric simulation, while the symbols (filled and empty cicles, squares and
triangles) are 3D data on the same timeslices from a Model $S_1$ star simulation on a $192^3$ grid with 
$\Delta x = \Delta y = \Delta z = 0.15$ resolution. Good agreement is observed.} 
 \renewcommand{\arraystretch}{0.75}
 \renewcommand{\topfraction}{0.6}
  \label{trK3D}
\end{center} 
\end{figure}

\begin{figure}
\begin{center}
\leavevmode
\epsfxsize=250pt
\epsfbox{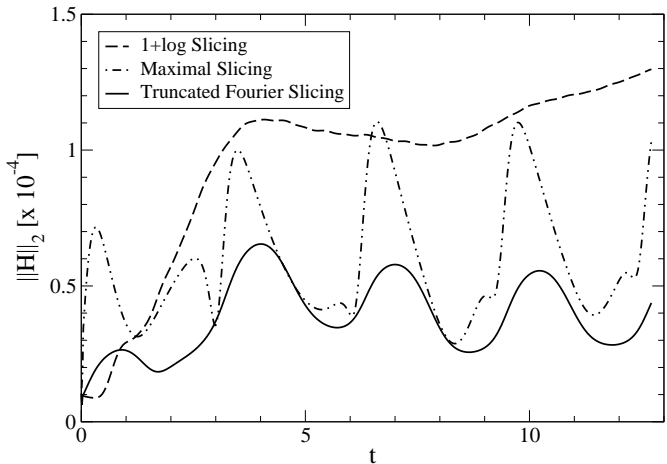}
\caption{ The L2-norm of the Hamiltonian is displayed as a function of time for 
three slicing conditions: (1) 1+log slicing, (2) Maximal Slicing and (3) Truncated 
Fourier Slicing. The latter is clearly smaller and leads to more stable simulations.}
 \renewcommand{\arraystretch}{0.75}
 \renewcommand{\topfraction}{0.6}
  \label{hamComparison}
\label{Fig7}
\end{center} 
\end{figure}

\begin{figure}
\begin{center}
\leavevmode
\epsfxsize=250pt
\epsfbox{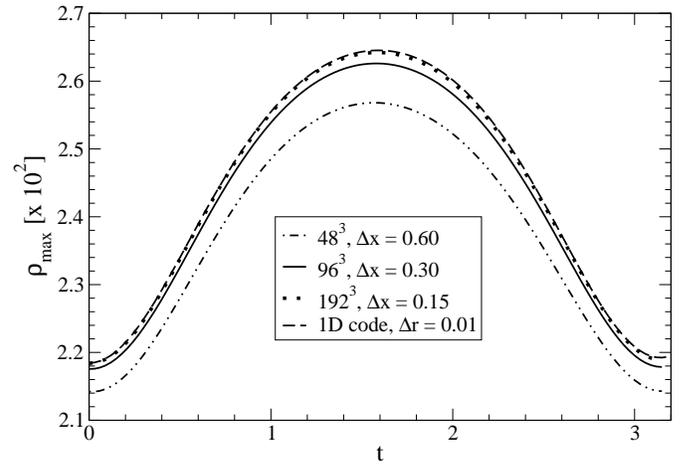}
\newline\newline\newline\newline
\epsfxsize=250pt
\epsfbox{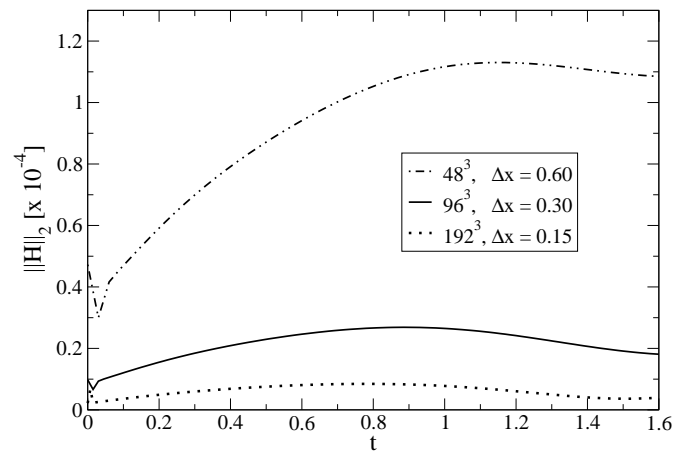}
\caption{(a) The maximum density $\rho$ of the Model $S_1$ soliton star is 
shown converging to the spherically symmetric 1D result. (b) The 
L2-norm of the Hamiltonian constraint is displayed for the three different resolutions.  The 
code is observed to exhibit roughly second order convergence.}
 \renewcommand{\arraystretch}{0.75}
 \renewcommand{\topfraction}{0.6}
  \label{hamConvergence}
\end{center} 
\end{figure}

\begin{figure}
\begin{center}
\leavevmode
\epsfxsize=250pt
\epsfbox{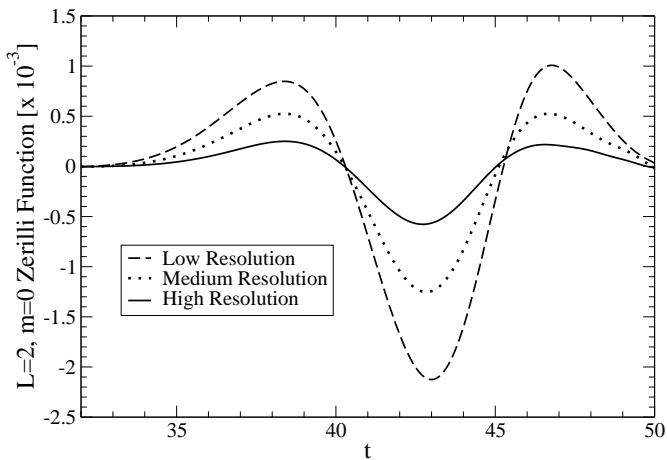}
\caption{The $\ell = 2, m = 0$ Zerilli function extracted at $r=43$ is shown for a perturbed Model $S_1$ solition star  at different resolutions labeled as: High for $\Delta x = \Delta y =0.196, \; \Delta z = 0.201$ with $n_x = n_y = 244, \; n_z = 238$,  Medium for  $\Delta x = \Delta y =0.294, \; \Delta z = 0.3015$
with $n_x = n_y = 164, \; n_z = 160$, and low  resolution for 
$\Delta x = \Delta y =0.392, \; \Delta z = 0.402 $ with $n_x = n_y = 122, \; n_z = 119$.  
The perturbation in this case is the asymmetric resolution itself and amplitude of the waveform is seen to converge away. The convergence rate is roughly second order. }
 \renewcommand{\arraystretch}{0.75}
 \renewcommand{\topfraction}{0.6}
  \label{waves}
\end{center}
\end{figure}

\begin{figure}
\begin{center}
\leavevmode
\epsfxsize=250pt
\epsfbox{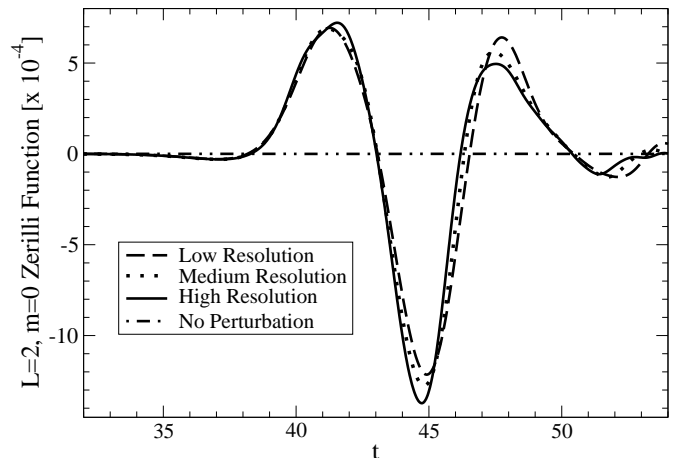}
\newline 
\newline 
\newline
\newline
\epsfxsize=250pt
\epsfbox{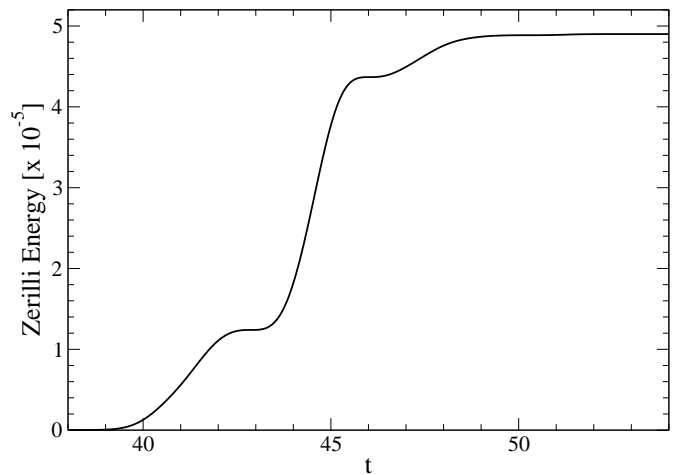}
\caption{(a)  The $\ell = 2, m = 0$ Zerilli function extracted at $r = 45$ is shown for a Model $S_1$ solition star under a $Y_{20}$ perturbation at  three different resolutions: High for  $\Delta x = \Delta y =\Delta z  = 0.196$ with $n_x = n_y = n_z =  244$,  Medium for  $\Delta x = \Delta y = \Delta z = 0.294$
with $n_x = n_y = n_z = 164$, and low  resolution for $\Delta x = \Delta y = \Delta z = 0.392$ 
with $n_x = n_y = n_z = 122$. The waveforms have roughly the same frequency and amplitude. 
(b) The Zerilli energy is displayed and is observed to flatted out on the same timescale 
as the gravitational wave signal.}
 \renewcommand{\arraystretch}{0.75}
 \renewcommand{\topfraction}{0.6}
  \label{waveformY20}
\end{center}
\end{figure}

\begin{figure}
\begin{center}
\leavevmode
\epsfxsize=250pt
\epsfbox{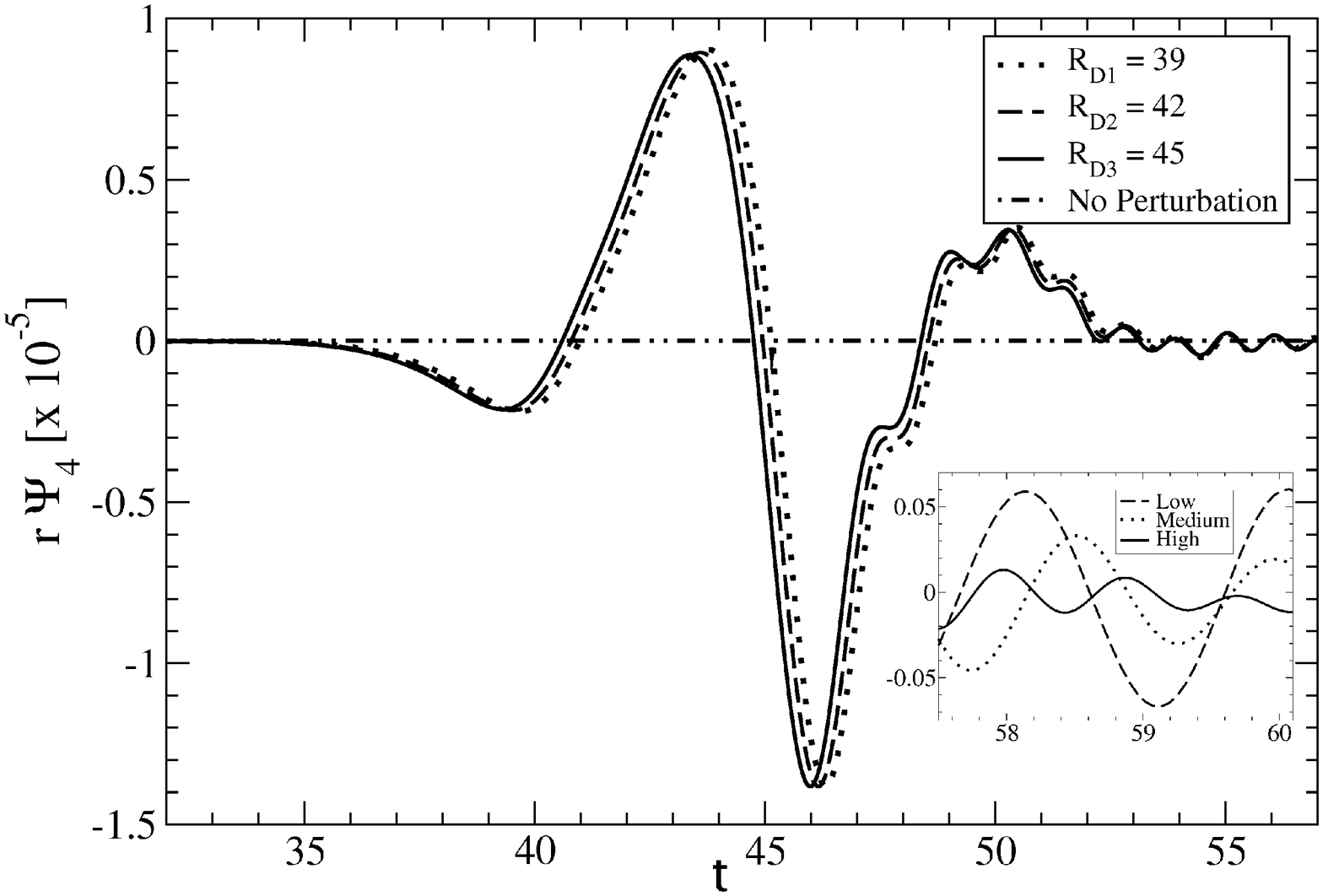}
\caption{
The Newman-Penrose scalar $\Psi_4$  multiplied by the radius at the 
detector is shown at three different detectors. The signals have been 
timeshifted by the distance in flat space between detectors to match 
the last detector. The agreement in both frequency and amplitude is good. 
The inset shows that the high frequency oscillations are noise and 
converge away with resolution.
}
 \renewcommand{\arraystretch}{0.75}
 \renewcommand{\topfraction}{0.6}
  \label{psi4}
\end{center}
\end{figure}
Fig.\ \ref{trK3D} shows $K$ as a function of $r$ on different time slices for 
a Model $S_1$ soliton star simulated on a $192^3$ three-dimensional grid with a resolution of 
$\Delta x = \Delta y = \Delta z = 0.15$.   
This is compared with the 1D result for a $\Delta r = 0.01$ resolution. 
The spherically symmetric 1D evolution of $K $ for a soliton star eigenstate is a periodic
oscillation between an upper bounding profile at approximately $t \approx \pi/4$ and 
a lower bounding profile at $t \approx 3 \pi/4$.
The oscillations of $K$ are observed to match the expected time dependence 
of Eq.\ (\ref{orderK}) and the agreement between the 1D and 3D results is good, 
indicating the truncated Fourier slicing condition is enforced accurately.

In order to demonstrate the advantages of applying this gauge condition, we
simulate the same Model $S_1$ star on a $96^3$ grid
with $\Delta x = \Delta y = \Delta z = 0.30$ with truncated Fourier slicing
and two widely used slicing conditions: maximal slicing ($K=0$) and 1+log 
slicing ($\partial_t \alpha = - 2 \alpha K$). 
Fig.\ \ref{Fig7} compares the L2-norm
of the hamiltonian constraint for these three simulations.
It can be seen that the Hamiltonian for the truncated Fourier slicing is by far 
the lowest (smaller by approximately a factor of $2$ than 1+log and Maximal slicing) and behaves is a stable manner.  These three test runs are carried out to a time of $ 4\pi$; 
for longer simulations that study perturbed soliton stars
we find that the use of the truncated Fourier slicing condition is necessary to 
control the coordinate drifting of metric functions to an acceptable tolerance 
and to perform the gravitational wave extraction accurately. 

A further test of the accuracy of the evolution code using the truncated Fourier slicing condition is 
the convergence test shown in Fig.\ \ref{hamConvergence}.
Fig.\ \ref{hamConvergence}(a) displays the first oscillation of the 
maximum of the density $\rho$ of the Model $S_1$ star 
converging to the 1D result as the 3D resolution is improved. 
We take three different resolutions: $48^3$ grid 
with $\Delta x = \Delta y = \Delta z = 0.60$ resolution, $96^3$ grid with  
$\Delta x = \Delta y = \Delta z = 0.30$ 
and $192^3$ grid with $\Delta x = \Delta y = \Delta z = 0.15$ resolution. 
These are compared to a high resolution spherically symmetric simulation ($\Delta r = 0.01$) 
with the 1D code. 
Fig.\ \ref{hamConvergence}(b) shows the convergence of the hamiltonian constraint, which is observed 
to be approximately second order.

\subsection{Nonradial Perturbations of Soliton Stars and Waveform Extraction}
\label{waveforms}

The focus of this section is the study of small nonspherical perturbations 
applied to spherically symmetric soliton stars
and the extraction of gravitational waveforms for such systems. 
We first perturb the star by imposing an asymmetry in the grid resolution.  We simulate
the Model $S_1$ star using a different resolution in the z-direction than 
in the x and y directions, while keeping the distance from the origin
to the boundary on each axis the same.
Fig.\ \ref{waves} shows the Zerilli waveform extracted at $r = 43 M_{\rm Pl}^2/m$ for three different resolutions labeled as 
(1) high resolution for 
$\Delta x = \Delta y =0.196, \; \Delta z = 0.201$ with
$n_x = n_y = 244, \; n_z = 238$, 
(2) medium resolution for  $\Delta x = \Delta y =0.294, \; \Delta z = 0.3015$
with $n_x = n_y = 164, \; n_z = 160$, and 
(3) low  resolution for $\Delta x = \Delta y =0.392, \; \Delta z = 0.402 $ with
$n_x = n_y = 122, \; n_z = 119$.  
The waveforms are observed to have the same frequency but different amplitudes.
For this series of simulations the uneven resolution is itself the perturbation,
and thus the waveform should converge away as the resolution is improved. 
We observe the appropriate second order converge rate, with a factor of four 
difference between the waveform amplitudes of the high resolution and the low resolution case.
Each of the simulations is performed using the truncated Fourier slicing condition 
and is run up to a time $ t \approx 60$ .

We next study an oscillating soliton star with a perturbation applied to the 
radial metric component $ g_{rr}$ of the form 
\begin{equation}
\delta g_{rr} (r,\theta,\phi) = \epsilon_{20} f(r/R) Y_{20}(\theta,\phi) g_{rr}(r)
\label{grrperteq}
\end{equation}
where $\epsilon_{20}$ is a constant much less than unity and the function
$f(r/R) = (r/R_p)^2 \exp(-c r^2)$. The constants $c$ and $R_p$ are chosen to localize 
the perturbation on the star and also not perturb the metric functions at the origin. 
This perturbation  does not significantly alter the mass of the star.  We expect that physical 
perturbations of soliton stars can be easily described
as linear combinations of spherical harmonics. This type of perturbation in the metric may mimic
the effect that gravitation of another star has on the oscillaton. 
In the current study we simply apply the perturbation of Eq.\ (\ref{grrperteq}) without 
resolving the initial data problem.  This procedure causes the constraints to no longer be obeyed
and hence makes the perturbation nonphysical.
A more complete treatment that incorporates an initial value problem solver that couples a 
scalar field and metric perturbation is the subject of future work.

We study the effects of a perturbation with $\epsilon_{20} = 10^{-4}$, $c = 1/16$ and $R_p =5$ 
on a Model $S_1$ soliton star.  In Fig.\ \ref{waveformY20}(a) the 
$\ell = 2, m = 0$ Zerilli functions extracted at a detector radius of $r =45 M_{\rm Pl}^2/m$
 are shown for three different resolutions. The waveforms oscillate through a 
couple of nodes and then damp to zero on a fairly short timescale. This is similar
to the case of boson stars \cite{futamase,us} and consistent
with the expectation that a star with a scalar field extending to infinity
would have only rapidly damped modes \cite{toymodel,futamase,kokkotas}.
The energy emitted in gravitational waves (see Fig.\ \ref{waveformY20}(b))
 is 
calculated and found to flatten out at about $5 \times 10^{-5} M_{\rm{Pl}}^2/m$.  
This suggests that the whole gravitational waveform has been extracted. 

We now investigate gravitational waveforms using the Newman-Penrose scalar $\Psi_4$
for the simulation described above.
 Fig.\ \ref{psi4} shows $r \Psi_4$ at three different detectors. The waveforms 
are timeshifted by the distance in flat space between each detector and the last detector 
to facilitate comparison. The frequencies and amplitudes of the waveforms at the three
detectors are in good agreement. This is consistent with the expected 
behavior of $\Psi_4 \propto 1/r$ in the far wave zone \cite{NPRefs}. 
High frequency oscillations appear at late times. They are likely 
an artifact of the numerical discretization. The frequency and amplitude of these oscillations 
are resolution dependent (see inset); these data show the amplitude to converge away to 
second order.  A more advanced technique such as adaptive mesh refinement that provides 
higher resolution would enable us to extend the convergence test to finer grid resolutions. 

Within all our 3D simulations we use the L2-norm of the Hamiltonian constrant as
a measure of the error. This is typically of the order of a few $\times 10^{-5}$ and
below $1 \times 10^{-4}$ for all the simulations in this section.

\section{Conclusions}
For the first time excited state soliton stars have been studied. Excited state configurations have field configurations characterized by nodes. For a given central field density $\phi_1(0)$, these stars are typically larger (greater radius) and more massive than corresponding ground state stars. We find that their mass profile is similar to that of ground state configurations; they have two branches separated at the maximum mass. The branch to the left of this maximum is traditionally called the $S$-branch because ground state configurations on this branch are stable to radial perturbations in the sense that they migrate to new configurations on the same branch. In spite of this similarity in the mass profile, we find that all excited state configurations are inherently unstable. Unlike ground state configurations, excite state stars on the $S$-branch do not migrate to new configurations on the excited state $S$-branch when perturbed.  They can either migrate to the ground state if they lose enough mass through scalar radiation or collapse to black holes. 

Higher excited state configurations migrate to the ground state sometimes 
cascading through intermediate excited states. We show one such migration 
for a five-node excited state star and find the four-node intermediate state 
that it is  nearest to during the migration process. Studying the stability 
of excited states is very important because they  may be intermediate 
states during the formation process of soliton stars.

Also, for the first time, the numerical evolution of ground state oscillaton stars is conducted on a full  3D grid, allowing the study of nonradial perturbations with gravitation waveform extraction. These simulations are challenging because soliton stars are very dynamic, with no equilibrium configurations having static metric components.  In this paper we compare slicing conditions previously used for 3D boson star simulations with a new slicing condition specifically designed for soliton stars. We find that the latter is more accurate and more stable. We see the energy density in 3D converging with resolution to that in 1D for the spherically symmetric case and the highest resolution 3D simulation matching the 1D result very accurately.

In 3D we explore two types of nonradial axisymmetric perturbations. First, the spherically symmetric 
1D eigenstate data is interpolated on the 3D grid with an imposed asymmetry 
in the grid resolution $\Delta x = \Delta y \ne \Delta z$. No explicit 
perturbation is applied. This has the advantage that  the initial data 
automatically satisfies the Einstein-Klein-Gordon equations. However, 
this perturbation is not easy to interpret physically. We successfully extract the 
Zerilli gravitational waveforms for a series of simulations at different 
resolutions and observe the amplitude of the signal due to this resolution-based
perturbation to converge away with improved 
resolution. The convergence rate is approximately second order as expected.
A second perturbation is applied explicitly to the metric function $g_{rr}$ and is 
proportional to the $Y_{20}$ spherical harmonic. Both the $\ell =2, m = 0$ Zerilli 
function and the Newman Penrose scalar $\Psi_4$ are extracted. 
This perturbation more closely represents a physical disturbance of the star 
and leads to a finite signal that does not converge away as in the previous case. 
 The waveforms damp rapidly and thus the complete
gravitational waveform could be extracted on a relatively short time scale.
This is consistent with expectations from perturbation theory for stars that have
a scalar field extending to infinity such as soliton stars and boson stars \cite{futamase}.
We also measure the energy radiated in 
gravitational waves as a function of time and find that 
it flattens as the signal damps out. 

This is an exploratory investigation that is meant as a starting point for future work. 
Possible improvements include
using an adaptive mesh refinement technique for better resolution
and finding a more general Gauge condition
that would be applicable to strongly perturbed soliton stars. 
These would enable the study of more advanced scenarios involving 
real fields such as soliton star collisions.

\section*{Acknowledgments}
We make extensive use of the Cactus Computational Toolkit, and its infrastructure for solving 
Einstein's Equations.  The simulations have been performed on the NCSA Tungsten cluster under
computer allocation TG-MCA02N014.  
We use the bspline package from the GNU software library \cite{GSL} for interpolation of initial data and the PETSc solver \cite{PETSc} for solving elliptic equations. 
We especially want to thank Francisco S. Guzman for advice and encouragement at the  beginning of the project, without which this paper would not have been completed. We would also like to acknowledge Francisco for developing the 3D evolution routine for the scalar field. We thank Peter Diener for useful discussions and Ed Seidel and Wai-Mo Suen for allowing us to use their 1D scalar field evolution code. We are very grateful to Doina Costescu and Cornel Costescu for hospitality, support and discussions during R.B.'s stays in Champaign-Urbana. We have
received partial funding from NSF grants PHY-0652952 and AST-0606710, and the
Sofja Kovalevskaja Program from the Alexander Von Humboldt Foundation.


\begin{thebibliography}{Rubinsteinetal}
\bibitem{toymodel} Similar to the case of boson stars and neutron stars, one can construct a toy model 
for oscillatons in which the star and the spacetime are each represented by
a semi-infinite string fastened at one end. The two strings are connected via a massless spring. Since the strings extend to infinity the system has no means to store energy and allows only strongly damped modes. See Ref.\ \cite{futamase} for details on the boson star model and
Ref.\ \cite{kokkotas} for the description of the neutron star case.
\end{thebibliography}
\end{document}